\documentclass[a4paper]{article}
\usepackage{multicol}
\usepackage{blindtext}
\usepackage{amsfonts,amssymb,amsmath,amsthm,cite}
\usepackage{ucs} 
\usepackage[utf8x]{inputenc}
\usepackage{graphicx}
\usepackage[english]{babel}
\usepackage{slashed}
\usepackage{textcomp}
\usepackage{bm}
\usepackage{titlesec}
\usepackage{stackengine}

\usepackage{etoolbox}
\patchcmd{\thebibliography}{\section*}{\section}{}{}
\usepackage{titlesec}
\titleformat*{\section}{\large\bf}
\usepackage{adjustbox}
\usepackage{nccmath}
\usepackage{mathtools}
\usepackage{xcolor}
\usepackage{footnote}
\usepackage{hyperref}
\usepackage{calligra} 
\hypersetup{
	colorlinks,
	citecolor=red,
	filecolor=black,
	linkcolor=red,
	urlcolor=black
}
\makesavenoteenv{tabular}
\mathtoolsset{showonlyrefs}
\evensidemargin=\oddsidemargin
\newtheorem{theorem}{Theorem}

\begin{document}
\begin{center}
	\large{\textbf{Renormalization, cutoff, and gluing for a quartic model with boundary}}
\end{center}
\begin{center}
	\large{\textbf{A. V. Ivanov}}
\end{center}
\begin{center}
	St. Petersburg Department
	of Steklov Mathematical Institute
	of Russian Academy of Sciences,\\
	27 Fontanka, St. Petersburg, 191023, Russia
\end{center}
\begin{center}
	Saint Petersburg State University,\\ 	7/9 Universitetskaya Emb., St. Petersburg, 199034, Russia
\end{center}

\begin{center}
E-mail: regul1@mail.ru
\end{center}

\vspace{5mm}
\begin{multicols}{2}

\textbf{Abstract.} In this paper, we study the renormalization of a three-dimensional quartic model on a compact connected Riemannian manifold with a boundary. We show that, in addition to the standard shift of the mass parameter, it is necessary to take into account boundary contributions involving singular coefficients. This indicates the need to extend the classical action within the framework of standard renormalization theory. The requirements of consistency regarding the gluing of manifolds lead to renormalization of an operator on the boundary.

\vspace{2mm}


\section{Introduction}

Perturbative expansions for most quantum field models \cite{3,10} contain divergent quantities \cite{6}, which, after the introduction of regularization, are transformed into singular functions (singularities), see \cite{Gelfand-1964, Vladimirov-2002}. If the problem is considered on a Riemannian manifold without a boundary and, moreover, simply in a flat space, then it is possible to identify a class of so-called multiplicatively renormalizable models, see \cite{7,105}, in which singularities can be eliminated by suitably multiplying parameters and fields by special renormalization constants.

When adding a boundary, the problem becomes more complicated, since the Green's function may have parts, see \cite{VZ}, that behave smoothly away from the boundary, but become singular when approaching the boundary. It should be noted that the very formulation of the problem in domains with boundary within the framework of standard quantum field theory is not new and has been addressed earlier, for example, see works \cite{grg-1,grg-2}. Also such situations arise, for example, in the framework of the functorial quantum field theory \cite{sk-3,sk-4,sk-5,sk-7} and are further complicated by the requirement of consistency in the process of introducing regularization and renormalization with respect to the gluing of manifolds and corresponding partition functions, see \cite{sk-14,sk-16}. With this formulation, standard regularization options do not work, for example, the dimensional one \cite{19,555} or the regularization with higher covariant derivatives \cite{Bakeyev-Slavnov,29-st}. At the same time, the class of quantum field models studied is rather narrow and actually limited to the two-dimensional case.

In this paper, we consider a three-dimensional quartic model \cite{gg-1} on a connected Riemannian manifold with a boundary. Although this model is super-renormalizable in the standard formulation \cite{29-3,Kh-2024}, a number of topical issues related to regularization, renormalization and singularities are currently being studied: a linearized case \cite{gg-2}, on the lattice \cite{gg-5}, as well as various aspects of constructing a probability measure \cite{gg-3,gg-4}. We also note recent work \cite{grg-3}, devoted to the problem of several $\delta$-potentials concentrated on surfaces, which can be obtained by shifting a boundary operator when gluing manifolds.

A special feature of the presented work is the use of a new special regularization by averaging \cite{Iv-2024,Ivanov-2022,ya-10} on a manifold with a boundary. It was shown earlier that the deformation of the theory by introducing such regularization is consistent with the gluing process, see \cite{sksk}. Nevertheless, the issues of renormalizability of models for $d>2$ were not considered in the proposed context, therefore, this paper provides the first example. The choice of the three-dimensional case is due to the nontriviality in terms of a set of singularities (power-law and logarithmic), while the model itself is much simpler than cases with a higher dimension.

The main results contain an explicit calculation of new counterterms both in the bulk and on the boundary, a discussion of extending the classical action, as well as a study of consistency with the gluing process. The work is structured as follows. Section \ref{sec:g1} discusses the problem statement, describes the quantum action and the regularization. In Section \ref{sec:g2}, the results are formulated, the issues of renormalizability of the model and consistency with the gluing process are discussed. Sections \ref{sec:g3} and \ref{sec:g4} present proofs, discussions of the results obtained, and open questions.

\section{Problem statement}
\label{sec:g1}
\textbf{Manifold.} Consider a three-dimensional smooth real compact connected orientable Riemannian manifold $(\mathcal{M},\mathrm{g})$, see \cite{Nakahara:2003nw}. Its boundary $\Sigma\neq\emptyset$ is smooth, closed, and can have a finite number of connected components whose distance is greater than a certain $t_b>0$. Next, we use the symbols $x,y$ to denote the elements of $\mathcal{M}$, which also serve as local coordinates. This simplification does not cause any confusion. The metric tensor and its determinant are denoted by $g_{\mu\nu}(x)$ and $g(x)$, and the geodesic distance between points is denoted by $r(\cdot,\cdot)$. In addition, we assume the presence of a product structure near the boundary: there exists $\mu>0$ such that 
\begin{equation*}
\{x\in\mathcal{M}:\,r(x,\Sigma)\leqslant1/\mu\}=\Sigma\times[0,1/\mu].
\end{equation*}
In this case, the metric near the boundary is labeled with the index $\Sigma$, and the corresponding coordinates are denoted by $u,v$. For example, the point of the manifold near the boundary has the form $x=(u,x^3)$, where $u=(u^1,u^2)$. Note that the product structure guarantees the form of the metric $\mathrm{g}=\mathrm{diag}(\mathrm{g}_\Sigma,1)$ in $\Sigma\times[0,1/\mu]$, where $\mathrm{g}_\Sigma$ is independent of $x^3$. This condition is stronger than the simple choice of a collar neighborhood, see section 9 in \cite{gg-1-1}, in which the block $\mathrm{g}_\Sigma$ may generally depend on the coordinate $x^3$.

\vspace{1mm}

\noindent\textbf{Classical action.} Let us choose $n\in\mathbb{N}$ and consider two vector functions: on the boundary $\eta\in C^\infty(\Sigma,\mathbb{R}^n)$ and in the bulk $\phi\in C^\infty(\mathcal{M},\mathbb{R}^n)$. In this case, we assume that $\phi_a|_{\Sigma}=\eta_a$, where the subscript indicates a separate component. Let us define a number of functionals
\begin{align*}
S_0[\phi]&=\frac{1}{2}\int_{\mathcal{M}}\mathrm{d}^3x\,g^{1/2}
g_{\mu\nu}(\partial^\mu\phi_a)(\partial^\nu\phi_a),\\
S_k[\phi]&=\frac{1}{k!}\int_{\mathcal{M}}\mathrm{d}^3x\,g^{1/2}t_k^{a_1\ldots a_k}
\phi_{a_1}\cdot\ldots\cdot\phi_{a_k}.
\end{align*}
Here, all coefficients are completely symmetric, and the repeating indices imply the corresponding summation. Let $t_2 = \mathrm{diag} \{m_a^2 \}$, where $m_a$ plays the role of a mass parameter for component $\phi_a$. Then, we define the classical action for a three-dimensional quartic model as the linear combination
\begin{equation}\label{g-32}
S_{\mathrm{cl}}[\phi]=S[\phi]+\hbar S_4[\phi],
\end{equation}
where the quadratic form is given by the equality
\begin{equation*}
S[\phi]=S_0[\phi]+S_2[\phi].
\end{equation*}
\textbf{Operators.} Consider the matrix-valued diagonal Laplace operator
\begin{equation*}
A^{ab}(x)=-\delta^{ab}g^{-1/2}(x)\partial_{x^\mu} g^{\mu\nu}(x)g^{1/2}(x)\partial_{x^\nu}+t_2^{ab}.
\end{equation*}
Then the function $\phi$ is uniquely representable as the sum of a background field $b$ and a fluctuation $\varphi$ satisfying the relations
\begin{equation*}
\varphi,b\in C^\infty(\mathcal{M},\mathbb{R}^n),\,\,\,
\varphi|_\Sigma=0,\,\,\, b|_\Sigma=\eta,
\end{equation*}
and also $A^{ab}b_b=0$ inside $\mathcal{M}$.
A direct substitution verifies the equality
\begin{equation*}
S[\phi]=S[b]+S[\varphi].
\end{equation*}
Next, the symbol $G_{bc}(x,y)$ denotes the Green's function in local coordinates for the operator $A^{ab}$ on $\mathcal{M}$ with zero Dirichlet conditions. It is clear that such a matrix is also diagonal. The corresponding spectral problem for such an operator is well-posed and studied in sufficient detail, see section 8 in \cite{sk-b-5}.

\vspace{1mm}

\noindent\textbf{Regularization.}
Let us choose $\Lambda>\mu$ and define the truncated manifold $\mathcal{M}_\Lambda$ by the relation
\begin{equation*}
\mathcal{M}_\Lambda=\{x\in\mathcal{M}:\,r(x,\Sigma)\geqslant 1/\Lambda\}.
\end{equation*}
The parameter $\Lambda$ is referred to as the regularizing one. Let the function $w\in C^\infty(\mathbb{R}_+,\mathbb{R}_+)$ satisfy two conditions:
\begin{equation*}
\mathrm{supp}(w)\subset[0,1/2],\,\,\,
\int_{0}^{1/2}\mathrm{d}s\,s^2w(s)=\frac{1}{4\pi}.
\end{equation*}
Then the averaging operator $\mathrm{H}_{x}^\Lambda$, acting from $C^\infty(\mathcal{M},\mathbb{R})$ to $C^\infty(\mathcal{M}_\Lambda,\mathbb{R})$ with respect to the variable "$x$", can be defined by the equality
\begin{equation}\label{g-2}
\mathrm{H}_{x}^\Lambda\phi_a(x)=\Lambda^3\int_{\mathcal{M}}\mathrm{d}^3y\,g^{1/2}w\big(\Lambda r(x,y)\big)\phi_a
\end{equation}
for all $x\in\mathcal{M}_\Lambda$. For convenience, the latter function is denoted by $\phi_a^\Lambda(x)$, and the symbol $\phi^\Lambda(x)$ represents the corresponding vector function. Taking into account all the above, the regularization of the classical action means the transition $S_{\mathrm{cl}}^{\phantom{1}}[\phi]\to S_{\mathrm{cl}}^\Lambda[\phi]$, which is performed by replacing
\begin{equation*}
S_4^{\phantom{1}}[\phi]\to S_4^\Lambda[\phi]\equiv S_4^{\phantom{1}}[\phi^\Lambda]\big|_{\mathcal{M}\to\mathcal{M}_\Lambda}.
\end{equation*}
Next, we assume that $\Lambda\gg\mu$, and we also require $1<2 t_b\Lambda$ for convenience, so that the number of connected components of the boundary for $\mathcal{M}_{\Lambda}$ and $\mathcal{M}$ is the same.

\vspace{1mm}

\noindent\textbf{Quantum action.} Let us define auxiliary functionals
\begin{align*}
	\Gamma_{k}[\varphi]=(k!)^{-1}\partial_s^kS_4^\Lambda[s\varphi+b]\big|_{s=0},
\end{align*}
where $k\in\{1,\ldots,4\}$. By construction, $\Gamma_k$ is proportional to the $k$-th degree of fluctuation $\varphi_a$, so each functional can be associated with a vertex with $k$ external lines, that is
\begin{center}
\begin{tabular}{ll}
$\Gamma_1\sim{\centering\adjincludegraphics[width = 0.9 cm, valign=c]{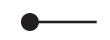}}$\,\,\,\,\,\,\,,
&
$\Gamma_2\sim{\centering\adjincludegraphics[width = 1.3 cm, valign=c]{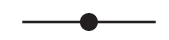}}$,
\\
$\Gamma_3\sim{\centering\adjincludegraphics[width = 1.3 cm, valign=c]{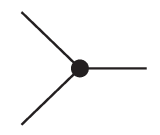}}$,
&
$\Gamma_4\sim{\centering\adjincludegraphics[width = 1.2 cm, valign=c]{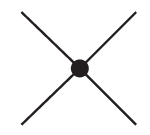}}$\,\,.
\end{tabular}
\end{center}
These lines correspond to the fields $\varphi_a$. Next, we define the operator $\mathbb{H}_0^{\mathrm{c(sc)}}$, which assigns a linear combination of connected (strongly connected) diagrams to each set of vertices $\Gamma$, which in total has an even number of external lines, by connecting the external lines in pairs in all possible ways using substitutions
\begin{equation*}
\varphi_a(x)\varphi_b(y)\to G_{ab}(x,y)
\end{equation*}
according to Wick's theorem, see \cite{Bog-R}. In this case, $\mathbb{H}_0^{\mathrm{c}}(\Gamma)=0$ if the number of external lines is odd, and also $\mathbb{H}_0^{\mathrm{c}}(1)=0$ for convenience. Let us note that a diagram is strongly connected if it remains connected after cutting any one internal line. Then a regularized quantum action is defined as a formal series of the form
\begin{equation}\label{g-1}
W^\Lambda_{\mathrm{reg}}[b]=S_{\mathrm{cl}}^\Lambda[b]-\mathbb{H}_0^{\mathrm{c}}\bigg[\exp\bigg(-\hbar\sum_{k=1}^4\Gamma_{k}[\varphi]\bigg)\bigg].
\end{equation}
The second term is a formal series in powers of the small parameter $\hbar$. Moreover, each order contains a finite number of terms. Symbolically, such a decomposition is usually written as a functional (path) "integral"
\begin{equation*}
e^{-W^\Lambda_{\mathrm{reg}}[b]}=\int\mathcal{D}\varphi\,e^{-S_{\mathrm{cl}}^\Lambda[b+\varphi]}.
\end{equation*}

\noindent\textbf{Renormalization.} As is known, see \cite{6} for reference, each coefficient in the expansion of the regularized quantum action \eqref{g-1} contains singular terms with respect to the parameter $\Lambda$. It turns out that in a flat space without a boundary, such objects can be eliminated by scaling parameters and fields, as well as adding a common normalization factor. This process is called multiplicative renormalization.

For example, in the space $\mathbb{R}^3$, the quartic model is super-renormalizable \cite{29-3}, since the auxiliary factors are series (with respect to $\hbar$) with a finite number of non-zero coefficients. As a result, the procedure reduces to shifting the mass coefficient function $t_2^{ab}\to t_2^{ab}+\Delta^{ab}$ and adding the formal constant $C$ to $S_{\mathrm{cl}}^\Lambda$, which is independent of the background field and contains infrared divergences. Within the framework of the proposed regularization, such values have been calculated in \cite{Kh-2024}, with the shift equal to
\begin{equation}\label{g-6}
\Delta^{ab}=\hbar\,\Delta^{ab}_1+\hbar^2\Delta^{ab}_2,
\end{equation}
where
\begin{align}\label{g-7}
\Delta^{ab}_1&=-\frac{1}{2}\Lambda\alpha t_4^{abcc}
,\\\label{g-8}
\Delta^{ab}_2&=\frac{L}{96\pi^2}t_4^{ac_1c_2c_3}t_4^{bc_1c_2c_3}.
\end{align}
Here, the value $\sigma>0$ is fixed and necessary to make combinations dimensionless, $L=\ln(\Lambda/\sigma)$, and the auxiliary constant $\alpha$ is
\begin{equation*}
\alpha=\int_{\mathbb{R}^3}\mathrm{d}^3x\int_{\mathbb{R}^3}\mathrm{d}^3y\,\frac{w(|x|)w(|y|)}{4\pi|x+y|}.
\end{equation*}
and is the value at the origin of the regularized free Green's function in flat space for $\Lambda=1$
\begin{equation}\label{g-14}
	\eta_\Lambda(z)=\int_{\mathbb{R}^{3\times2}}
	\frac{\mathrm{d}^3x\mathrm{d}^3y\,w(|x|)w(|y|)}{4\pi|z+(x+y)/\Lambda|}.
\end{equation}
Theorem \ref{g-t1} in Section \ref{sec:g2} provides results on generalizing the renormalization process to the case of $\mathcal{M}$, including a description of new boundary counterterms and coefficients that are actually a consequence of adding the boundary.

\vspace{1mm}

\noindent\textbf{Gluing of manifolds.} Let us cut the manifold $\mathcal{M}$ into two submanifolds $\mathcal{M}_l$ and $\mathcal{M}_r$ along some surface $Y$ of codimension 1. We assume that the new parts satisfy all the requirements formulated in Section \ref{sec:g1}. In this case, the boundary $\Sigma$ splits into two disconnected parts: $\Sigma=Y_l\cup Y_r$, where
\begin{equation*}
\partial\mathcal{M}_i=Y_i\cup Y\,\,\,\mbox{for}\,\,\,i=l,r.
\end{equation*} 
Further, each submanifold has its own classical action
\begin{equation*}
S_{\mathrm{cl}}[\phi_i;\mathcal{M}_i],\,\,\,\mbox{where}\,\,\,
\phi_i\in C^\infty(\mathcal{M}_i,\mathbb{R}^n),
\end{equation*}
satisfying the conditions
\begin{equation*}
\phi_i\big|_{Y_i}=\eta_i\in C^\infty(Y_i,\mathbb{R}^n),\,\,\,\phi_i\big|_{Y}=\psi\in C^\infty(Y,\mathbb{R}^n).
\end{equation*}
An additional subscript is introduced here. Since it consists of $\{l,r\}$, there is no confusion with the numbering of the individual components. At the same time $S_{\mathrm{cl}}[\phi;\mathcal{M}]=S_{\mathrm{cl}}[\phi_l;\mathcal{M}_l]+S_{\mathrm{cl}}[\phi_r;\mathcal{M}_r]$, where $\phi=\phi_l+\phi_r$. Note that $\phi_l$ and $\phi_r$ in the latter equality should be understood in the sense of embedding into $C^\infty(\mathcal{M},\mathbb{R}^n)$ using the operation of extension by zero. Repeating the regularization procedure, hereinafter the third argument denotes the truncated manifold $\mathcal{M}_\Lambda$ or $\mathcal{M}_{i,\Lambda}$, which plays the role of an integration domain in the interaction term. Next, the regularized quantum actions are constructed as follows
\begin{equation*}
W_{\mathrm{reg}}^\Lambda[b_i;\mathcal{M}_i,\mathcal{M}_{i,\Lambda}],
\end{equation*}
where $b_i$ is the corresponding background field, uniquely constructed from the boundary values. In Theorem 1 of \cite{sksk}, it was shown that the functional integration of the combination
\begin{equation*}
\exp\bigg(-\sum_{i=l,r} W_{\mathrm{reg}}^\Lambda[b_i;\mathcal{M}_i,\mathcal{M}_{i,\Lambda}]\bigg)
\end{equation*}
with respect to the boundary function $\psi$, which must be understood in the sense of Gaussian averaging of formal series in powers of the Planck's constant $\hbar$ with respect to the Dirichlet-to-Neumann operator (or the Poincaré--Steklov map, see \cite{gg-1-2}) in quadratic form for the boundary field $\psi$, is equal to 
\begin{equation*}
\exp\Big(- W_{\mathrm{reg}}^\Lambda[b;\mathcal{M},\mathcal{M}_{l,\Lambda}\cup\mathcal{M}_{r,\Lambda}]\Big).
\end{equation*}
And after eliminating the "scar" with the help of the limit transition $\mathcal{M}_{l,\Lambda}\cup\mathcal{M}_{r,\Lambda}\longrightarrow\mathcal{M}_{\Lambda}$, which consists in returning the truncated parts (by continuous extension) of the manifolds near the boundary $Y$, it turns out
\begin{equation*}
	\exp\Big(- W_{\mathrm{reg}}^\Lambda[b;\mathcal{M},\mathcal{M}_{\Lambda}]\Big).
\end{equation*}
Note that the limit transition is well defined and continuous, because the manifolds appear only in integration domains, while all integrands are finite. Such a transition from the quantum action (partition functions) on submanifolds to the action on the whole manifold is commonly called gluing. At the same time, it was proved (see \cite{sksk}) that such a gluing procedure is compatible with the process of multiplicative renormalization. Below, in Theorem \ref{g-t2} in Section \ref{sec:g2}, a generalization is proved and it is shown that additional boundary counterterms arise on a manifold with boundary, which are absent in the framework of the standard multiplicative procedure. They must be cancelled additionally, thus renormalizing the Dirichlet-to-Neumann operator on the boundary (or the integration measure).

\section{Results}
\label{sec:g2}
\noindent\textbf{Renormalized action.} We define several auxiliary functions: two "mass" quadratic terms
\begin{equation*}
	\hat{S}_{2,i}^\Lambda[\phi]=\frac{1}{2}\int_{\mathcal{M}_\Lambda}\mathrm{d}^3x\,g^{1/2}\Delta^{ab}_i
	\phi_{a}^\Lambda \phi_{b}^\Lambda\,\,\,\mbox{for}\,\,\,i=1,2,
\end{equation*}
see \eqref{g-6}, the boundary term
\begin{align*}
	S_\Sigma[\eta]=\frac{L}{32}\int_{\Sigma}\mathrm{d}^2u\,g^{1/2}_{\Sigma}\eta_{a}^{\phantom{1}} \eta_{b}^{\phantom{1}}t_4^{abcc},
\end{align*}
where $L=\ln(\Lambda/\sigma)$, as well as
\begin{equation*}
	\hat{\Gamma}_{k,i}[\varphi]=(k!)^{-1}\partial_s^k\hat{S}_{2,i}^\Lambda[s\varphi+b]\big|_{s=0},
\end{equation*}
where $k,i\in\{1,2\}$. The last functionals correspond to vertices with one and two external lines 
\begin{equation*}
\Gamma_{1,i}\sim{\centering\adjincludegraphics[width = 1.4 cm, valign=c]{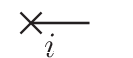}},\,\,\,
\Gamma_{2,i}\sim{\centering\adjincludegraphics[width = 1.9 cm, valign=c]{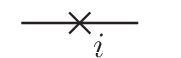}}.
\end{equation*}
Additionally, we define the constant 
\begin{equation*}
C=C_{\mathrm{in}}+C_{\mathrm{b}},
\end{equation*}
by dividing it into a bulk (internal) part and a boundary part. At the same time
\begin{align}\label{g-34}
C_{\mathrm{in}}&=\hbar\,c_{\mathrm{in},1}+\hbar^2c_{\mathrm{in},2}+\hbar^3c_{\mathrm{in},3},\\\label{g-35}
C_{\mathrm{b}}&=\hbar\,c_{\mathrm{b},1}+\hbar^2c_{\mathrm{b},2},
\end{align}
where the coefficients are determined by the equalities
\begin{fleqn}
\begin{align}
c_{\mathrm{in},1}&=\mathrm{Vol}(\mathcal{M}_\Lambda)\Lambda^2\alpha^2t^{aabb}/8,\\\label{g-21}
c_{\mathrm{in},2}&=\mathrm{Vol}(\mathcal{M}_\Lambda)t_4^{abce}t_4^{abce}
\\&\,\,\,\,\,\,\,\,\,\,\,\,\,\,\,\,\,\,\,\,\,\,\,\times\frac{\Lambda}{2^43}\bigg(\int_{\mathbb{R}^3}\mathrm{d}^3z\,\eta_1^4(z)-\frac{L\alpha}{4\pi^2}\bigg),\\
c_{\mathrm{in},3}&=\mathrm{Vol}(\mathcal{M}_\Lambda)L
t_4^{abce}t_4^{cedg}t_4^{dgab}/(2^{14}3\pi^2),\\
c_{\mathrm{b},1}&=\mathrm{Vol}(\Sigma)(-1)\Lambda t_4^{aabb}/(2^9\pi^2),\\\label{g-29}
c_{\mathrm{b},2}&=\mathrm{Vol}(\Sigma)t^{abce}_4t_4^{abce}
\\
&\,\,\,\,\,\,\,\,\,\,\,\,\,\,\,\,\,\,\,\,\,\,\,\times \frac{L}{2^43\pi}\bigg(\frac{\theta}{2^7\pi^2}-\int_{|z|<1}\mathrm{d}^3z\,\eta_1^3(z)\bigg).
\end{align}
\end{fleqn}
Here the number $\theta\approx0.629615\pm10^{-6}$ does not depend on any parameters and is the numerical result for the integral from Section \ref{sec:g3}. After adding new terms to the classical action, the regularized action \eqref{g-1} takes the following form
\begin{multline}\label{g-5}
S_{\mathrm{cl}}^\Lambda[b]+\hbar\hat{S}_{2,1}^\Lambda[b]+\hbar^2\hat{S}_{2,2}^\Lambda[b]+\hbar S_\Sigma[\eta]+C
	\\
	-\mathbb{H}_0^{\mathrm{c}}\bigg[\exp\bigg(-\sum_{k,i=1,2}\hbar^i\hat{\Gamma}_{k,i}-\hbar\sum_{k=1}^4\Gamma_{k}\bigg)\bigg],
\end{multline}
We further denote it by $W^\Lambda_{\mathrm{ren}}[b]$, and call it the renormalized action. Note that countervertices have two parts proportional to $\hbar$ and $\hbar^2$. Therefore, the formal series in powers of the Planck's constant has a finite number of terms at each order after decomposition.

\begin{theorem}\label{g-t1}
Taking into account all the above, the renormalized action \eqref{g-5} in each order of $\hbar$ does not contain singular contributions, and the corresponding coefficients are finite when the regularization is removed.
\end{theorem}

\begin{theorem}\label{g-t2} Let the notations of the last part of Section \ref{sec:g1} be in force. Taking into account the result of Theorem \ref{g-t1}, by analogy with \eqref{g-5}, we construct the renormalized action
\begin{equation*}
	W_{\mathrm{ren}}^\Lambda[b_i;\mathcal{M}_i,\mathcal{M}_{i,\Lambda}],
\end{equation*}
on each submanifold $\mathcal{M}_l$ and $\mathcal{M}_r$. Next, we define a constant on the boundary
\begin{equation*}
C_Y=C_{\mathrm{b}}\mathrm{Vol}(Y)/\mathrm{Vol}(\Sigma),
\end{equation*}
as well as the counterterm on the boundary
\begin{align*}
	S_Y[\psi]=\frac{L}{32}\int_{Y}\mathrm{d}^2u\,g^{1/2}_{Y}\psi_{a}^{\phantom{1}} \psi_{b}^{\phantom{1}}t_4^{abcc}.
\end{align*}
Then the functional integration of the quantity 
\begin{equation*}
	\exp\bigg(-\sum_{i=l,r} W_{\mathrm{ren}}^\Lambda[b_i;\mathcal{M}_i,\mathcal{M}_{i,\Lambda}]\bigg)
\end{equation*}
with respect to the field $\psi$ with the weight
\begin{equation}\label{g-33}
	\exp\Big(2\hbar S_Y[\psi]+2C_Y\Big)
\end{equation}
in the sense of Gaussian averaging of formal series in powers of the Planck's constant $\hbar$ with respect to the Dirichlet-to-Neumann operator on the boundary $Y$ in quadratic form for the boundary field $\psi$ is equal to
\begin{equation}\label{g-36}
	\exp\Big(- W_{\mathrm{ren}}^\Lambda[b;\mathcal{M},\mathcal{M}_{l,\Lambda}\cup\mathcal{M}_{r,\Lambda}]\Big)
\end{equation}
and after the removal of the "scar" by the limit transition
\begin{equation}\label{g-37}
\mathcal{M}_{l,\Lambda}\cup\mathcal{M}_{r,\Lambda}\to\mathcal{M}_{\Lambda}
\end{equation}
is transformed into action
\begin{equation}\label{g-38}
\exp(-W_{\mathrm{ren}}^\Lambda[b;\mathcal{M},\mathcal{M}_{\Lambda}]).
\end{equation}
\end{theorem}

\noindent\textbf{Remark 1.} Note that in the case of a manifold with boundary, the usual multiplicative renormalization is not sufficient, since singularities appear localized on the boundary. Such objects are cancelled by adding the counterterms $S_\Sigma[\eta]+C_{\mathrm{b}}$. From the point of view of the standard paradigm, the quartic model \eqref{g-32} on the manifold with boundary is non-renormalizable, and the problem is solved by extending the classical action
\begin{equation*}
	S_{\mathrm{cl}}[\phi]\to S_{\mathrm{cl}}[\phi]+\int_{\Sigma}\mathrm{d}^2u\,g^{1/2}_{\Sigma}\eta_{a}^{\phantom{1}} \eta_{b}^{\phantom{1}}t^{ab}_{\mathrm{b}},
\end{equation*}
where $t^{ab}_{\mathrm{b}}$ is a symmetric coefficient of dimension $[x^{-1}]$. In this formulation, adding the constant $C_{\mathrm{b}}$ is responsible for shifting the action at the boundary, and adding $S_\Sigma[\eta]$ should be understood as shifting the coupling constant.
\begin{equation*}
	t^{ab}_{\mathrm{b}}\to t^{ab}_{\mathrm{b}}+\hbar Lt_4^{abcc}/32.
\end{equation*}
In this case, the renormalization process becomes multiplicative. However, the extended action does not fit into the conditions of the theorem from \cite{sksk}, since the "boundary" term is actually a bulk one with a singular density localized on the boundary. Therefore, renormalization of the Dirichlet-to-Neumann operator, formulated in Theorem \ref{g-t2}, is additionally necessary. Furthermore, we note that it is possible to introduce its own coefficient on each disconnected piece of the boundary.

\vspace{1mm}

\noindent\textbf{Remark 2.} Note that the shift parameters
\begin{equation*}
C_{\mathrm{in}}\sim\mathrm{Vol}(\mathcal{M}_\Lambda),\,\,\,
C_{\mathrm{b}}\sim\mathrm{Vol}(\Sigma).
\end{equation*}
At the same time, they do not contain any other dependence on the manifolds. If we consider individual coefficients, then $c_{\mathrm{in},1}$ and $c_{\mathrm{in},2}$ contain power-law singularities and depend on the kernel of the averaging operator \eqref{g-2}, whereas $c_{\mathrm{in},3}$ is proportional to $L$ and does not depend on the regularization properties. At the boundary, the situation is different: the power-law factor $c_{\mathrm{b},1}$ does not depend on the regularization, while the logarithmic one $c_{\mathrm{b},2}$ depends.

\vspace{1mm}

\noindent\textbf{Remark 3.} Since the coefficient $C_{\mathrm{in}}$ is proportional to the volume of the integration domain from the interaction functional, the quantum actions from \eqref{g-36} and \eqref{g-38} have different constants proportional to 
\begin{equation*}
C_{\mathrm{in}}\frac{\mathrm{Vol}(\mathcal{M}_{l,\Lambda}\cup\mathcal{M}_{r,\Lambda})}{\mathrm{Vol}(\mathcal{M}_{\Lambda})}
\,\,\,\mbox{and}\,\,\,
C_{\mathrm{in}},
\end{equation*}
respectively. This fact, in particular, guarantees the finiteness of the coefficients of the formal series not only for the final case without the scar, but also at intermediate stages.

\vspace{1mm}

\noindent\textbf{Remark 4.} Let us supplement the classical action for the quartic model \eqref{g-32} with the cubic term
\begin{equation*}
S_{\mathrm{cl}}[\phi]\to S_{\mathrm{cl}}[\phi]+\hbar^{1/2}S_3[\phi],
\end{equation*}
the regularization of which can be introduced similarly by smoothing the fields $\phi\to\phi^\Lambda$ and truncating the integration domain $\mathcal{M}\to\mathcal{M}_\Lambda$. In this case, the counterterms described in Theorem \ref{g-t1} are not sufficient to renormalize the quantum action. They need to be supplemented with the following changes. First, we need to add the bulk linear part
\begin{multline*}
\frac{\hbar^{1/2}}{2}\int_{\mathcal{M}_\Lambda}\mathrm{d}^3x\,g^{1/2}\phi_{a}^\Lambda\\\times
\bigg[-\Lambda\alpha t_3^{acc}+\frac{\hbar L}{48\pi^2}t_4^{ac_1c_2c_3}t_3^{c_1c_2c_3}\bigg].
\end{multline*}
Secondly, it is necessary to add the boundary linear part
\begin{align*}
\frac{\hbar^{1/2}L}{16}\int_{\Sigma}\mathrm{d}^2u\,g^{1/2}_{\Sigma}\eta_{a}^{\phantom{1}}t_3^{acc},
\end{align*}
and, thirdly, additionally change one volume constant as follows
\begin{equation*}
c_{\mathrm{in},1}\to c_{\mathrm{in},1}+\frac{L\mathrm{Vol}(\mathcal{M}_\Lambda)}{192\pi^2}t_3^{c_1c_2c_3}t_3^{c_1c_2c_3}.
\end{equation*}
The appearance of linear terms actually entails some difficulties, since in this case, the standard understanding of renormalization requires an extension of the classical action with a linear term $\hbar^{-1/2}S_1[\phi]$, which contains a negative power of Planck's constant and, thus, generates an unlimited number of terms in each order. Nevertheless, the purely formal standard singularities are cancelled even in this case.

\vspace{1mm}

\noindent\textbf{Remark 5.} The definition for the regularized quantum action \eqref{g-1} is based on the decomposition of the functional \eqref{g-32} near the classical solution of the free linear equation. It is for this reason that additional difficulties have arisen with adding a linear term, described in Remark 4. It is worth noting that it is possible to decompose near the solution of the classical equation with the interaction. In this case, the results of Theorem \ref{g-t1} concerning renormalization issues remain true, and the linear term can be added without any difficulties. Moreover, the background field does not have to be "small", and the classical action takes on a more familiar form
\begin{equation*}
S[b+\varphi]+\hbar S_4[b+\varphi]\to\frac{S[\hat{b}+\sqrt{\hbar}\varphi]+ S_4[\hat{b}+\sqrt{\hbar}\varphi]}{\hbar^2}.
\end{equation*}
Here $\hat{b}=\sqrt{\hbar}b$ and $\hat{b}+\sqrt{\hbar}\varphi$ represents a deviation from the classical solution by a small fluctuation. However, it is important to note that in this case the proof of Theorem \ref{g-t2} loses its relevance, since in the process of gluing of manifolds it was important to choose the free solution, see \cite{sksk}. The fact is that it is not entirely clear whether, for the nonlinear case, the chosen extremal trajectory on the entire manifold is a gluing of extremal trajectories on submanifolds, and the question of their number remains open as well.

\section{Proof}
\label{sec:g3}
\noindent\textbf{Green's functions.} Next, let $\sigma>0$ be such that $\mu<\sigma\sigma_1/(\sigma_1+\sigma)$, where $\sigma_1=\sigma\Lambda/(\sigma+\Lambda)$, and also for all $c>0$ and $y\in\mathcal{M}$, we define a geodesic ball
\begin{equation*}
\mathrm{B}_{c}(y)=\{x\in\mathcal{M}:\,r(x,y)\leqslant c\}.
\end{equation*}
It is known from general theory, see, for reference, Section 8 in \cite{sk-b-5}, that the Green's function $G_{ab}(x,y)$ is a symmetric function, and the expansion near the diagonal $(x\sim y)$ for it can be written in two ways. First, if $y\in\mathcal{M}_{\sigma_1}$ and $x\in\mathrm{B}_{1/\sigma}(y)$, then
\begin{equation}\label{g-3}
G_{ab}(x,y)=\frac{\delta^{ab}}{4\pi r(x,y)}+p_{0ab}(x,y),
\end{equation}
see the decomposition near the diagonal in \cite{15,sk-b-11}. In this case, the second term is a continuous bounded function. Now let us take $y\in\Sigma\times[0,1/\sigma_1]$, then $y=(v,s)$, where $v\in\Sigma$. Let us assume additionally 
\begin{equation*}
(u,t)=x\in\mathrm{B}_{1/\sigma}(y)\subset\Sigma\times[0,1/\mu],	
\end{equation*}
then $r^2(x,y)=r^2_\Sigma(u,v)+(s-t)^2$. In addition, we define the function $\hat{r}^2(x,y)=r^2_\Sigma(u,v)+(s+t)^2$,
then the asymptotic expansion has the form
\begin{equation}\label{g-4}
	G_{ab}(x,y)=\frac{\delta^{ab}}{4\pi r(x,y)}-\frac{\delta^{ab}}{4\pi \hat{r}(x,y)}+p_{1ab}(x,y).
\end{equation}
In this case, the third term is continuous and bounded and vanishes at $s=0$ and/or $t=0$ for $x\neq y$.

\vspace{1mm}

\noindent\textbf{Regularized functions.} After the introduction of regularization in the classical action, the fields $\phi$ are replaced by smoothed functions $\phi^\Lambda$, see \eqref{g-2}. With such a substitution, using Wick's theorem in constructing the series \eqref{g-1} is equivalent to replacing Green's functions
\begin{equation*}
G_{ab}(x,y)\to G_{ab}^\Lambda(x,y)\equiv\mathrm{H}_{x}^\Lambda\mathrm{H}_{y}^\Lambda G_{ab}(x,y)
\end{equation*}
for a similar series in which the fluctuations $\varphi$ are not averaged. Thus, the task of calculating singular contributions is reduced to working with the deformed Green's functions. Let us define two auxiliary functions
\begin{align}\label{g-10}
\rho_\Lambda(x,y)&=\mathrm{H}_{x}^\Lambda\mathrm{H}_{y}^\Lambda\big(4\pi r(x,y)\big)^{-1},\\
\hat{\rho}_\Lambda(x,y)&=\mathrm{H}_{x}^\Lambda\mathrm{H}_{y}^\Lambda\big(4\pi \hat{r}(x,y)\big)^{-1},
\end{align}
then, denoting the smoothed versions of $p_{0}$ and $p_{1}$ by the superscript $\Lambda$, we have similar equalities
\begin{equation}\label{g-9}
	G^\Lambda=\rho_\Lambda+p_{0}^\Lambda\,\,\,\mbox{and}\,\,\,G^\Lambda=\rho_\Lambda-\hat{\rho}_\Lambda+p_{1}^\Lambda,
\end{equation}
see for comparison \eqref{g-3} and \eqref{g-4}. Next, we recall the definition of \eqref{g-14} for $\hat{\eta}_\Lambda(\cdot)$. Such a function is the main approximation of $\rho_\Lambda(x,y)$ for large $\Lambda$, if we understand $z$ as the value of $x-y$ in the flat space. Let additionally $z_1=(u,s)$ and $z_2=(v,t)$, where $s,t\geqslant0$ and $u,v\in\mathbb{R}^2$, as well as $r=|u-v|$, then we define
\begin{equation}\label{g-25}
g_\Lambda(r,s-t)=\hat{\eta}_\Lambda(z_1-z_2).
\end{equation} 
In this case, the function $g_\Lambda(r,s+t)$ is the main approximation of $\hat{\rho}_\Lambda(x,y)$ for $\Lambda\to+\infty$, if we understand $z_1,z_2$ as the local coordinates of $x,y$ in the flat space. If we choose $z=((0,0),2s)$, where $s\geqslant0$, then we can define the function
\begin{equation}\label{g-15}
	\hat{\eta}(s)=\hat{\eta}_1(z),
\end{equation}
which, in the combination of $\Lambda\hat{\eta}(s\Lambda)$, provides the main approximation for $\hat{\rho}_\Lambda(x,x)$ near the boundary when $(u,s)=x\in\Sigma\times [1/\Lambda,1/\sigma_1)$.

\vspace{1mm}

\noindent\textbf{Renormalization for parts $\sim\hbar$.} Let us check that the functional \eqref{g-5} does not contain singularities with respect to the parameter $\Lambda$ in the coefficient at $\hbar^1$. Let us write out all the terms
\begin{equation*}
\hat{S}_{2,1}^\Lambda[b]+S_\Sigma[\eta]+
\mathbb{H}_0^{\mathrm{c}}\Big(\hat{\Gamma}_{2,1}+\Gamma_{2}+\Gamma_{4}\Big)+c_{\mathrm{in},1}+c_{\mathrm{b},1}.
\end{equation*}
It is convenient to divide them into two groups. First, consider the combination  
\begin{equation}\label{g-11}
\mathbb{H}_0^{\mathrm{c}}(\Gamma_{2})+\hat{S}_{2,1}^\Lambda[b]+S_\Sigma[\eta],
\end{equation}
the first two terms in which are integrals over $\mathcal{M}_\Lambda$ with a total density of the form
\begin{equation*}
\mathbb{H}_0^{\mathrm{c}}(\Gamma_{2})+\hat{S}_{2,1}^\Lambda\sim
\frac{1}{4}b_{a}^\Lambda b_{b}^\Lambda\Big(t_4^{abce}
G_{ce}^\Lambda(x,x)+2\Delta_1^{ab}\Big).
\end{equation*}
In this case, it is convenient to divide the integration area into two parts: $\mathcal{M}_{\sigma_1}$ and $\Sigma\times[1/\Lambda,1/\sigma_1)$. Note that the function $\rho_\Lambda(x,x)$ is present in both areas, see \eqref{g-9}. Using the definition of \eqref{g-10} and the transition to normal coordinates, see Section $\mathrm{II}.8$ in \cite{gg-1-3}, we obtain
\begin{equation}\label{g-12}
\rho_\Lambda(x,x)=\Lambda\alpha+\mathcal{O}(1/\Lambda)\,\,\,\mbox{when}\,\,\,\Lambda\to+\infty
\end{equation}
uniformly by $x\in\mathcal{M}_\Lambda$. Thus, the singular part of the function $t_4^{abcc}\rho_\Lambda$ cancels the coefficient $\Delta_1^{ab}$ completely. Considering the boundedness of $p_0^\Lambda$ and $p_1^\Lambda$, we see that the singular component of the combination $\mathbb{H}_0^{\mathrm{c}}(\Gamma_{2})+\hat{S}_{2,1}^\Lambda$ can only be near the boundary in an integral of the form
\begin{equation*}
-\frac{1}{4}\int_{1/\Lambda}^{1/\sigma_1}\mathrm{d}s\int_{\Sigma}\mathrm{d}^2u\,g_\Sigma^{1/2}b_{a}^\Lambda b_{b}^\Lambda t_4^{abcc}\hat{\rho}_\Lambda(x,x).
\end{equation*}
Here $x=(u,s)$. Next, we simplify the integral by replacing the field $b^\Lambda$ with the main part $\hat{\eta}(u)$, and instead of the function $\hat{\rho}_\Lambda(x,x)$ by choosing $\hat{\rho}(x,x)$. The latter substitution is possible because the domain of the first integral is bounded away from zero and starts with $1/\Lambda$. Considering that $\hat{\rho}(x,x)=1/(8\pi s)$, we obtain the singular part in the form
\begin{equation*}
	-\frac{\ln(\Lambda/\sigma)}{32}\int_{\Sigma}\mathrm{d}^2u\,g_\Sigma^{1/2}\eta_{a} \eta_{b}t_4^{abcc}=-S_\Sigma[\eta].
\end{equation*}
Thus, the singular part of $\mathbb{H}_0^{\mathrm{c}}(\Gamma_{2})+\hat{S}_{2,1}^\Lambda$ in the second region near the boundary compensates for the singularity from the term $S_\Sigma[\eta]$. Therefore, combination \eqref{g-11} is finite when the regularization is removed.

\vspace{1mm}

\noindent The second combination is the sum of the diagrams
\begin{equation*}
\mathbb{H}_0^{\mathrm{c}}\Big(\hat{\Gamma}_{2,1}+\Gamma_{4}\Big)+c_{\mathrm{in},1}+c_{\mathrm{b},1}.
\end{equation*}
The first two terms are integrals over $\mathcal{M}_\Lambda$. Let us use \eqref{g-7}, then after regrouping the density is written as
\begin{multline*}
\frac{1}{8}t_4^{abce}G_{ab}^\Lambda G_{ce}^\Lambda+\frac{1}{2}G_{ab}^\Lambda\Delta^{ab}_1=-\frac{\Lambda^2\alpha^2}{8}t^{aabb}\\+
\frac{1}{8}t_4^{abce}\big(G_{ab}^\Lambda-\Lambda\alpha\delta_{ab}\big)\big(G_{ce}^\Lambda-\Lambda\alpha\delta_{ce}\big).
\end{multline*}
Here, all the Green's functions are taken on the diagonal. It is clear that after adding the integral, the first term is cancelled by the constant $c_{\mathrm{in},1}$. Consider the second line and split the integral into two parts: $\mathcal{M}_{\sigma_1}$ and $\Sigma\times[1/\Lambda,1/\sigma_1)$. In the first case, taking into account \eqref{g-12}, the singularity $\Lambda\alpha$ completely cancels the similar contribution in the first expansion from \eqref{g-9}. The remaining density does not contain any other singularities, so the integral over $\mathcal{M}_{\sigma_1}$ is finite. In the second case, there is additionally the function $\hat{\rho}_\Lambda$, which is singular near the boundary. Considering the behavior of the function $p_{1}^\Lambda$ to zero when approaching the boundary, only the following part of the form remains
\begin{equation*}
\frac{t_4^{aabb}}{8}\int_{1/\Lambda}^{1/\sigma_1}\mathrm{d}s\int_\Sigma\mathrm{d}^2u\,g_\Sigma^{1/2}
\Big(\hat{\rho}_\Lambda(x,x)\Big)^2,
\end{equation*}
where $x=(u,s)$. Taking into account the definition of \eqref{g-14}, for $\Lambda\to+\infty$ we get the singularity
\begin{equation}\label{g-16}
\frac{t_4^{aabb}}{8}\int_{1/\Lambda}^{1/\sigma_1}\mathrm{d}s\int_\Sigma\mathrm{d}^2u\,g_\Sigma^{1/2}
\Big(\Lambda\hat{\eta}(s\Lambda)\Big)^2,
\end{equation}
because for the function $\hat{\rho}_\Lambda$ on the diagonal in the domain $\Sigma\times[1/\Lambda,1/\sigma_1)$ decomposition of the form
\begin{equation}\label{g-22}
\hat{\rho}_\Lambda(x,x)=\Lambda\hat{\eta}(s\Lambda)+\mathcal{O}(1/\Lambda)
\end{equation}
is valid, see formulas \eqref{g-14} and \eqref{g-15}. Next, we note, see \cite{Iv-2024,ya-10}, that the function $\hat{\eta}_1(z)=(4\pi|z|)^{-1}$ in the region $|z|\geqslant1$, so $\hat{\eta}(s)=1/(8\pi s)$ for all $s\geqslant1/2$. Moving to the limit, we see that the constant in the main order \eqref{g-16} cancels $c_{\mathrm{b},1}$.

\vspace{1mm}

\noindent As an auxiliary consequence of this section, we note that combinations of the form
\begin{equation}\label{g-13}
\mathbb{H}_1^{\mathrm{c}}\big(\Gamma_{3}\big)+\hat{\Gamma}_{1,1}\,\,\,\mbox{and}\,\,\,
\mathbb{H}_2^{\mathrm{c}}\big(\Gamma_{4}\big)+\hat{\Gamma}_{2,1}
\end{equation}
do not contain any singularities. In the diagrammatic language, the relations can be written as
\begin{align}\label{g-30}
0&\stackrel{\mathrm{s.p.}}{=}3\raisebox{+.07\height}{\centering\adjincludegraphics[width = 1.4 cm, valign=c]{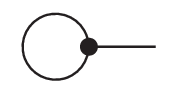}}\,+
\raisebox{-.1\height}{\centering\adjincludegraphics[width = 1.4 cm, valign=c]{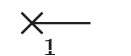}},\\
\label{g-31}
0&\stackrel{\mathrm{s.p.}}{=}
6{\centering\adjincludegraphics[width = 1.4 cm, valign=c]{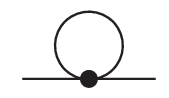}}+
\raisebox{-.1\height}{\centering\adjincludegraphics[width = 1.9 cm, valign=c]{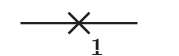}},
\end{align}
where the symbol $\stackrel{\mathrm{s.p.}}{=}$ indicates equality for singular components. Note that in the proof, it was important that the fluctuation fields at the boundary were equal to zero.

\vspace{1mm}

\noindent\textbf{Renormalization at $\sim\hbar^2$.} Let us write out all the parts of functional \eqref{g-5} proportional to $\hbar^2$
\begin{multline*}
\hat{S}_{2,2}^\Lambda[b]+c_{\mathrm{in},2}+c_{\mathrm{b},2}
+\mathbb{H}_0^{\mathrm{c}}\big(\hat{\Gamma}_{2,2}\big)
\\
-2^{-1}\mathbb{H}_0^{\mathrm{c}}\big(\hat{\Gamma}_{1,1}+\hat{\Gamma}_{2,1}+\Gamma_{1}+\Gamma_{2}+\Gamma_{3}+\Gamma_{4}\big)^2.
\end{multline*}
Reveal the brackets and remove the parts with an odd number of external lines. Next, we note that the combinations
\begin{align*}
&\mathbb{H}_0^{\mathrm{c}}\big(\Gamma_{1}^2\big),\,\,\,\mathbb{H}_0^{\mathrm{c}}\big(\Gamma_{1}^{\phantom{1}}\Gamma_{3}^{\phantom{1}}+\Gamma_{1}^{\phantom{1}}\hat{\Gamma}_{1,1}^{\phantom{1}}\big),
\\
&\mathbb{H}_0^{\mathrm{c}}\big(\hat{\Gamma}_{1,1}^2+2\hat{\Gamma}_{1,1}^{\phantom{1}}\Gamma_{3}^{\phantom{1}}+\big[\mathbb{H}_1^{\mathrm{c}}(\Gamma_{3}^{\phantom{1}})\big]^2\big),\\
&\mathbb{H}_0^{\mathrm{c}}\big(\Gamma_{2}^2\big),\,\,\,\mathbb{H}_0^{\mathrm{c}}\big(\Gamma_{2}^{\phantom{1}}\big[\hat{\Gamma}_{2,1}^{\phantom{1}}+\mathbb{H}_2^{\mathrm{c}}(\Gamma_{4}^{\phantom{1}})\big]\big),
\\
&\mathbb{H}_0^{\mathrm{c}}\big(\hat{\Gamma}_{2,1}^2+2\hat{\Gamma}_{2,1}^{\phantom{1}}\big[\mathbb{H}_2^{\mathrm{c}}(\Gamma_{4}^{\phantom{1}})\big]+\big[\mathbb{H}_2^{\mathrm{c}}(\Gamma_{4}^{\phantom{1}})\big]^2\big),
\end{align*}
do not contain singularities due to the previously performed renormalization of the parts of $\sim\hbar$, since the corresponding densities include combinations of \eqref{g-13}. Thus, considering the relation
\begin{align*}
\mathbb{H}_0^{\mathrm{c}}\big(\Gamma_3^2-\big[\mathbb{H}_1^{\mathrm{c}}(\Gamma_{3}^{\phantom{1}})\big]^2\big)&=3!{\centering\adjincludegraphics[width = 1.4 cm, valign=c]{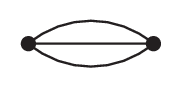}},
\\
\mathbb{H}_0^{\mathrm{c}}\big(\Gamma_4^2-\big[\mathbb{H}_1^{\mathrm{c}}(\Gamma_{4}^{\phantom{1}})\big]^2\big)&=4!{\centering\adjincludegraphics[width = 1.4 cm, valign=c]{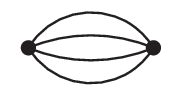}},
\end{align*}
only the following two diagram combinations
\begin{align}\label{g-17}
	-3&
	{\centering\adjincludegraphics[width = 1.4 cm, valign=c]{fig/skk-12.eps}}+\hat{S}_{2,2}^\Lambda[b]
	,\\\label{g-18}
	-12
&	{\centering\adjincludegraphics[width = 1.4 cm, valign=c]{fig/skk-11.eps}}+c_{\mathrm{in},2}+c_{\mathrm{b},2}
+{\centering\adjincludegraphics[width = 1.6 cm, valign=c]{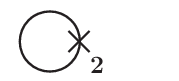}}
\end{align}
may contain singular contributions that need to be cancelled by a suitable selection of free coefficients $\Delta_2^{ab}$, $c_{\mathrm{in},2}$, and $c_{\mathrm{b},2}$. Considering the diagram from \eqref{g-17}, we note that a singular contribution can only be given by a combination in which all the Green's functions $G_{ab}^\Lambda$ are replaced by the main approximation $\delta_{ab}\rho_\Lambda$, both inside the volume and near the boundary. Moving to normal coordinates, we notice that the functions $\rho_\Lambda$, in turn, can be replaced by the main approximation $\hat{\eta}_\Lambda$ from \eqref{g-14}. Then, using the asymptotic behavior for the integral
\begin{equation*}
\int_{\mathrm{B}_{1/\sigma}(0)}\mathrm{d}^3z\,
\hat{\eta}_\Lambda^3(z)=\frac{\ln(\Lambda/\sigma)}{16\pi^2}+\mathcal{O}(1),
\end{equation*}
making sure that $\Delta_2^{ab}$ is equal to \eqref{g-8}. In addition, we note that the selection of such a coefficient leads to two consequences
\begin{align}\label{g-19}
	0&\stackrel{\mathrm{s.p.}}{=}-2^33\raisebox{+.09\height}{\centering\adjincludegraphics[width = 1.7 cm, valign=c]{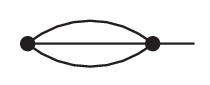}}\,+
	\raisebox{-.1\height}{\centering\adjincludegraphics[width = 1.2 cm, valign=c]{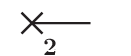}},
	\\\label{g-20}
	0&\stackrel{\mathrm{s.p.}}{=}-2^53\raisebox{+.09\height}{\centering\adjincludegraphics[width = 1.9 cm, valign=c]{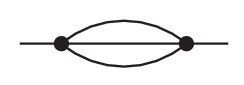}}+
	\raisebox{-.1\height}{\centering\adjincludegraphics[width = 1.7 cm, valign=c]{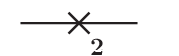}},
\end{align}
in which, again, the equality to zero of the fluctuations at the boundary should be taken into account. Moving on to the diagrams from \eqref{g-18}, let us break down the calculation into several steps.

\vspace{1mm}

\noindent1) Let us split the integration domains in both diagrams into two parts
\begin{align*}
{\centering\adjincludegraphics[width = 0.9 cm, valign=c]{fig/skk-11.eps}}\,\,\,\,&:\,\,\,
\mathcal{M}_\Lambda\times\mathcal{M}_{\sigma_1}\,\,\,\mbox{and}\,\,\,
\mathcal{M}_\Lambda\times\big(\Sigma\times[1/\Lambda,1/\sigma_1)\big),\\
{\centering\adjincludegraphics[width = 1.1 cm, valign=c]{fig/skk-18.eps}}&:\,\,\,
\mathcal{M}_{\sigma_1}\,\,\,\mbox{and}\,\,\,
\big(\Sigma\times[1/\Lambda,1/\sigma_1)\big).
\end{align*}
\noindent2) Let us consider the first domains. Substituting the first expansion from \eqref{g-9}, we verify that the singular part with $p_0^\Lambda$ is cancelled by $\Delta_2^{ab}$. Next, moving to normal coordinates, replacing the main parts of $\rho_\Lambda$ with $\hat{\eta}_\Lambda$ and using the asymptotic expansion for the integral
\begin{equation*}
\int_{\mathrm{B}_{1/\sigma}(0)}\mathrm{d}^3z\,\hat{\eta}_\Lambda^4(z)=
\Lambda\int_{\mathbb{R}^3}\mathrm{d}^3z\,\hat{\eta}_1^4(z)+\mathcal{O}(1),
\end{equation*}
taking into account \eqref{g-12}, we see that the singular parts for the sum of the first and fourth terms from \eqref{g-18} have the form
\begin{equation*}
\Lambda\mathrm{Vol}(\mathcal{M}_{\sigma_1})\frac{t_4^{abce}t_4^{abce}}{2^43}\bigg(\frac{L\alpha}{4\pi^2}-\int_{\mathbb{R}^3}\mathrm{d}^3z\,\hat{\eta}_1^4(z)\bigg).
\end{equation*}
This value is cancelled by a part of the coefficient $c_{\mathrm{in},2}$ from \eqref{g-21} after splitting $\mathrm{Vol}(\mathcal{M}_{\Lambda})$ to the sum of $\mathrm{Vol}(\mathcal{M}_{\sigma_1})$ and $\mathrm{Vol}(\Sigma)/\sigma$.

\vspace{1mm}

\noindent3) Consider the second domains. Using the decomposition near the boundary of \eqref{g-9}, we notice that the parts with $p_1^\Lambda$ are being cancelled again. Next, note that after switching to normal coordinates, the functions $\rho_\Lambda$ and $\hat{\rho}_\Lambda$ can be replaced by the main parts in flat local coordinates. Thus, the remaining singular part for the fourth term of \eqref{g-18} has the form
\begin{equation}\label{g-27}
	\mathrm{Vol}(\Sigma)\frac{\Delta_2^{aa}}{2}\bigg(\frac{\Lambda\alpha}{\sigma}-\frac{L}{8\pi}\bigg),
\end{equation}
where we have used \eqref{g-12} and \eqref{g-22}, the relation $\hat{\eta}(s)=1/(8\pi s)$ for all $s\geqslant1/2$, as well as the notation $L=\ln(\Lambda/\sigma)$. In turn, the singular contribution of the first part of \eqref{g-18} is contained in the integral
\begin{equation}\label{g-28}
-\frac{t^{abce}_4t_4^{abce}}{2^33}\mathrm{Vol}(\Sigma)
\mathrm{I}(\Lambda,\sigma)\pi,
\end{equation}
where
\begin{multline*}
\mathrm{I}(\Lambda,\sigma)=
	\int_{1/\Lambda}^{1/\sigma_1}\mathrm{d}t
	\int_{1/\Lambda}^{t+1/\sigma}\mathrm{d}s
	\int_0^1\mathrm{d}r\,r\times\\\times\Big(g_\Lambda(r,s-t)-g_\Lambda(r,s+t)\Big)^4.
\end{multline*}
For convenience, integration over a cylinder was used here, rather than integration over a ball. Let us perform several transformations. First, note that the region $[0,1]$ can be replaced by $\mathbb{R}_+$. Secondly, $\sigma_1$ can be replaced with $\sigma$. Then, after scaling the variables, we get
\begin{multline*}
	\mathrm{I}(\Lambda,\sigma)\stackrel{\mathrm{s.p.}}{=}
	\int_{1}^{\Lambda/\sigma}\mathrm{d}t
	\int_{1}^{t+\Lambda/\sigma}\mathrm{d}s
	\int_{\mathbb{R}_+}\mathrm{d}r\,r\times\\\times\Big(g_1(r,s-t)-g_1(r,s+t)\Big)^4.
\end{multline*}
One can study the asymptotics for such an integral by differentiation. Applying the operator $\Lambda\partial_\Lambda$ and performing "reverse" scaling, we get two terms
\begin{align*}
\dot{\mathrm{I}}_1&=\int_{1/\gamma}^{1}\mathrm{d}t
\int_{\mathbb{R}_+}\mathrm{d}r\,r\Big(g_\gamma(r,1)-g_\gamma(r,2t+1)\Big)^4,
\\
	\dot{\mathrm{I}}_2&=
	\int_{1/\gamma}^{2}\mathrm{d}s
	\int_{\mathbb{R}_+}\mathrm{d}r\,r\Big(g_\gamma(r,s-1)-g_\gamma(r,s+1)\Big)^4,
\end{align*}
where $\gamma=\Lambda/\sigma$. It is necessary to find asymptotic expansions for such integrals at $\gamma\to+\infty$ up to and including a constant. Let us first consider $\dot{\mathrm{I}}_1$. Note that it is possible to move to $\gamma\to+\infty$ in the lower limit of integration. Further, the arguments for the functions $g_\gamma(a,b)$ are such so that $a^2+b^2\geqslant1>1/\gamma$, therefore, taking into account the results of \cite{ya-10,sksk}, equality is valid in the specified region
\begin{equation}\label{g-24}
g_\gamma(a,b)=\frac{1}{4\pi(a^2+b^2)^{1/2}}.
\end{equation}
Substituting and calculating explicitly four two-dimensional integrals, we obtain $\dot{\mathrm{I}}_1=\theta_1/(4\pi)^4+o(1)$, where
\begin{align*}\label{g-23}
\theta_1&=\frac{2}{3}-2\ln(3)-\frac{3\pi^2}{8}-\frac{3}{2}\ln(2)\ln(3)
\\
&+\frac{3}{2}\Big(\mathrm{Li}_2(3)-\mathrm{Li}_2(-3)+i\pi\ln(3)\Big)\\
&\approx 0.015424\pm10^{-6}.
\end{align*}
In the case of $\dot{\mathrm{I}}_2$, we can proceed similarly: choose $\gamma\to+\infty$ in the integration limit and in the function
\begin{equation*}
\lim_{\gamma\to+\infty}g_\gamma(r,s+1)=\frac{1}{4\pi(r^2+(s+1)^2)^{1/2}}.
\end{equation*}
In the case of $g_\gamma(r,s-1)$, a transition to the limit is possible only if the degree of the function does not exceed two. Next, we note that 
\begin{equation}\label{g-26}
\dot{\mathrm{I}}_2=\mathrm{j}_1-\frac{\mathrm{j}_2}{2\pi}+\frac{\theta_2+5}{(4\pi)^4}+o(1).
\end{equation}
In this case, special functions and their asymptotic expansions are determined by the relations
\begin{fleqn}
\begin{align*}
\mathrm{j}_1=\int_{0}^{2}\mathrm{d}s
\int_{\mathbb{R}_+}&\mathrm{d}r\,rg_\gamma^4(r,s-1)\\&=2\int_{0}^{1}\mathrm{d}s
\int_{\mathbb{R}_+}\mathrm{d}r\,rg_\gamma^4(r,s)\\
&=\frac{\gamma}{2\pi}\int_{\mathbb{R}^3}\mathrm{d}^3z\,\hat{\eta}_1^4(z)-\frac{1}{(4\pi)^4},
\end{align*}
\end{fleqn}
where we used equalities \eqref{g-25} and \eqref{g-24},
\begin{fleqn}
\begin{align*}
\mathrm{j}_2=\int_{0}^{2}\mathrm{d}s
	\int_{\mathbb{R}_+}&\mathrm{d}r\,rg_\gamma^3(r,s-1)\\&=\frac{2}{(4\pi)^3}+\frac{1}{\pi}
	\int_{\mathrm{B}_1}\mathrm{d}^3z\,\hat{\eta}_\gamma^3(z)
	\\&=\frac{4\ln(\gamma)+2}{(4\pi)^3}+\frac{1}{\pi}
	\int_{\mathrm{B}_1}\mathrm{d}^3z\,\hat{\eta}_1^3(z).
\end{align*}
\end{fleqn}
The number $\theta_2$ is determined by relation \eqref{g-26} and is represented as an integral over $[0,2]\times\mathbb{R}_+$ with density in the form of the difference between the left and right sides, in which the limit $\gamma\to+\infty$ was taken. Such an integral does not depend on anything and can be obtained numerically
\begin{equation*}
\theta_2\approx0.61419\pm10^{-6}.
\end{equation*}
Thus, summing up all the parts and integrating with respect to the parameter $L=\ln(\Lambda/\sigma)$, we see that the following decomposition is valid
\begin{multline*}
	\mathrm{I}(\Lambda,\sigma)\stackrel{\mathrm{s.p.}}{=}
	\frac{\Lambda}{2\pi\sigma}\int_{\mathbb{R}^3}\mathrm{d}^3z\,\hat{\eta}_1^4(z)
	-\frac{4L^2}{(4\pi)^4}\\-\frac{L}{2\pi^2}\int_{\mathrm{B}_1}\mathrm{d}^3z\,\hat{\eta}_1^3(z)+\frac{L\theta}{(4\pi)^4},
\end{multline*}
where
\begin{equation*}
\theta=\theta_1+\theta_2\approx0.629615\pm10^{-6}.
\end{equation*}
Substituting the obtained asymptotic expansion into \eqref{g-28} and summing it with the rest of the parts from \eqref{g-18}, taking into account the answer of \eqref{g-27}, we see that the choice of the coefficient $c_{\mathrm{b},2}$ according to \eqref{g-29} cancels all remaining singularities. Thus, the finiteness of all terms of $\sim\hbar^2$ was clearly verified.

\vspace{1mm}

\noindent\textbf{Renormalization at $\sim\hbar^k$ for $k>2$.} Contributions to higher quantum corrections consist of connected diagrams that are constructed using four vertices $\Gamma_i$, where $i\in\{1,2,3,4\}$, and four countervertices $\hat{\Gamma}_{k,n}$, where $k,n\in\{1,2\}$. In this case, singularities can appear for three reasons, which follows from the applicability of the standard $\mathcal{R}$-operation \cite{33-rev7}. Firstly, because of the appearance of the Green's function on the diagonal, the so-called "tadpole". However, such combinations are cancelled by corresponding counterdiagrams with the coefficient $\Delta_1^{ab}$, which is a direct consequence of relations \eqref{g-30} and \eqref{g-31}. Secondly, singularities can appear due to the presence of "triple" lines, that is, three identical Green's functions. However, such combinations are also cancelled by counterdiagrams with the coefficient $\Delta_2^{ab}$, which is a direct consequence of relations \eqref{g-19} and \eqref{g-20}. The third and final possibility of a singularity is to obtain a "triple" line by integrating a "double" one. This option is feasible only in a diagram of the form
\begin{equation*}
D={\centering\adjincludegraphics[width = 1.4 cm, valign=c]{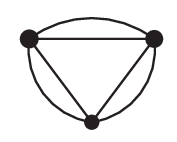}},
\end{equation*}
which is obtained from the connections of three quartic vertices and is contained in the correction $\sim\hbar^3$. Writing out the corresponding coefficient from \eqref{g-5}, we see that it is necessary to select $c_{\mathrm{in},3}$ in such a way that the equality holds 
\begin{equation*}
c_{\mathrm{in},3}-2^53^2D\stackrel{\mathrm{s.p.}}{=}0.
\end{equation*}
Considering the fact that the singularity is logarithmic, we can use the already known result for the flat space from \cite{Kh-25}. Thus, we get
\begin{equation*}
D\stackrel{\mathrm{s.p.}}{=}\frac{\mathrm{Vol}(\mathcal{M}_\Lambda)\pi^2L}{2^{11}3^3(4\pi)^4}t_4^{abce}t_4^{cedg}t_4^{dgab}
,
\end{equation*}
hence, the answer for the coefficient stated in the theorem follows. Thus, Theorem \ref{g-t1} is proved. In turn, the proof of Theorem \ref{g-t2} follows from the result of \cite{sksk} after cancelling the extra boundary terms \eqref{g-33}.

\section{Conclusion}
\label{sec:g4}

In this paper, a three-dimensional quartic model was renormalized on a connected Riemannian manifold with boundary, all renormalization constants were calculated, and new boundary terms were found that are absent in the framework of the usual model without boundary. The results are formulated in Theorem \ref{g-t1}. Then, the relation of gluing of manifolds and partition functions (quantum actions) was extended to the renormalized actions by adding a weight function to the functional integral, which actually renormalizes the Dirichlet-to-Neumann boundary operator. The results are given in Theorem \ref{g-t2}. Additionally, in the remarks after the theorems, the features of the obtained coefficients, the options for choosing a classical solution, the addition of odd terms to the classical action, and the renormalization process as a whole are discussed.

One of the interesting questions is the generalization of the properties of the manifold $\mathcal{M}$. In Section \ref{sec:g1}, an assumption was formulated about the product structure near the boundary. This fact has greatly simplified the calculation of boundary singularities. The author could not find a precise relaxation of this condition. It is expected that the main results of the theorems will remain valid. Cases with piecewise smooth boundaries are also not entirely clear.

Another important task is to calculate the singularities for the four-dimensional case. Due to the higher dimension, calculations become much more complicated. Moreover, a new dependence on the properties of the manifold may appear, both inside the volume (bulk) and near the boundary. For the flat case without a boundary, three-loop calculations have already been performed, see \cite{Iv-2024-1}, so it is expected that they should be preserved in the leading singularities.

\textbf{Acknowledgements.} This study is supported by the Ministry of Science and Higher Education of the Russian Federation, agreement number 075-15-2025-013. The author expresses gratitude to N.V.Kharuk and I.V.Korenev for useful comments, as well as to L. and K. for creating a positive working environment.

\end{multicols}

\end{document}